\documentclass[prl,twocolumn,showpacs,amsmath,amssymb]{revtex4}
\usepackage[latin1]{inputenc}
\usepackage{graphicx}
\usepackage{dcolumn}
\usepackage{bm}
\begin{document}
\title{Mixed state properties of superconducting  MgB$_2$ single crystals}
\author{M.~Zehetmayer}
\email{zehetm@ati.ac.at}
\author{M.~Eisterer}
\author{H.~W.~Weber}
\affiliation{Atominstitut der Österreichischen Universitäten,
A-1020 Vienna, Austria}
\author{J.~Jun}
\author{S.~M.~Kazakov}
\author{J.~Karpinski}
\affiliation{Solid State Physics Laboratory, ETH, CH-8093 Zürich,
Switzerland }
\author{A.~Wisniewski}
\affiliation{Institute of Physics, Polish Academy of Sciences,
PL-02-668 Warsaw, Poland}
\date{\today}
\begin{abstract}
We report on measurements of the magnetic moment in
superconducting MgB$_2$ single crystals. We find
$\mu_0H_{c2}^c(0)$ = 3.2~T, $\mu_0H_{c2}^{ab}(0)$ = 14.5~T,
$\gamma$ = 4.6, $\mu_0H_c(0)$ = 0.28~T, and $\kappa(T_c)$ = 4.7.
The standard Ginzburg-Landau and London model relations lead to a
consistent data set and indicate that MgB$_2$ is a clean limit
superconductor of intermediate coupling strength with very
pronounced anisotropy effects.
\end{abstract}

\pacs{74.25.Ha, 74.60.Ec, 74.70.Ad}
\maketitle

The recent discovery of superconductivity in MgB$_2$ \cite{Nag01}
has attracted a lot of attention. Especially the rather high
transition temperature of nearly 40~K in such a simple compound is
of interest for applications, but also for an analysis of the
physical mechanism leading to superconductivity. Several
experiments indicate a phonon mediated s - wave BCS mechanism
\cite{Bud01,Qui02}. Different models are proposed to explain the
particular properties of MgB$_2$ \cite{Haa01,Shu01}. Their
correctness has to be checked by experiments, but only a few
results are available on single crystals
\cite{Lee01,Pra01,Elt02,Man02,Ang01,Sol01,Wel02}.

We report in this Letter on magnetization measurements on single
crystalline MgB$_2$ in magnetic fields applied parallel and
perpendicular to the uniaxial crystallographic ($\equiv$ c) axis.
A detailed evaluation allows us to obtain the temperature
dependence of the most important reversible mixed state
parameters, such as the critical magnetic fields, the
characteristic lengths, the Ginzburg-Landau (GL) parameter and the
anisotropy. We will show that MgB$_2$ is a clean limit
superconductor of intermediate coupling strength with very
pronounced anisotropy effects.

\begin{figure}
  \centering \includegraphics[width = \columnwidth]{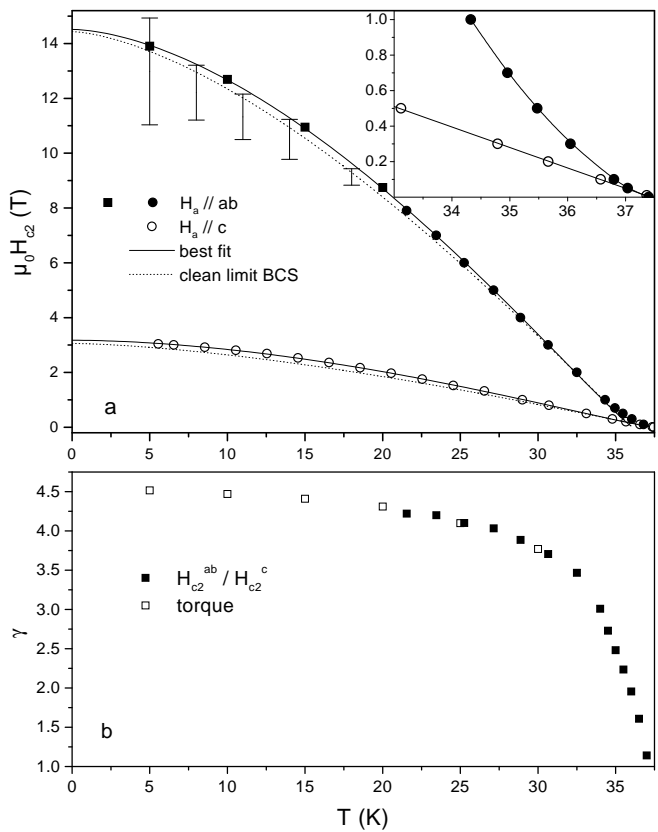}
\caption{(a) Upper critical field for $H_a \| c
 $ ($H_{c2}^c$) and $H_a \| ab$ ($H_{c2}^{ab}$).
  $H_{c2}^{ab}$ is obtained from: (i) a direct evaluation of
  $m(T)$ for $T >$ 21~K (solid circles) and (ii) $H_{c2}^{ab} = \gamma
  H_{c2}^c$ for $T <$ 21~K (solid squares). The error bars indicate
  the extrapolation uncertainties of the reversible moments measured
   up to 8~T. The BCS curves according to \cite{Hel66} are fitted to the experimental slope of
    $H_{c2}$ near $T_c$. (b) Anisotropy from SQUID ($H_{c2}^{ab} /
H_{c2}^c$) and torque measurements.} \label{fig-hc2}
\end{figure}

Several single crystals of MgB$_2$ were grown using high pressure
cubic anvils. Details of the process will be published elsewhere
\cite{Kar02}. Two crystals (sample A: $a × b × c \cong 660 × 570 ×
21~\mu m^3$; sample B: $a × b × c \cong 600 × 384 × 54~\mu m^3$)
were investigated by magnetic methods. The transition temperature
($T_c$) of each sample was obtained from the ac - susceptibility
measured in a 1~T quantum interference device (SQUID)
magnetometer. Sample A shows an onset of $T_c$ at 38~K and a
rather broad transition of about 1~K. A linear fit of $H_{c2}^c$
vs. $T$ near $T_c$ indicates a "bulk transition temperature" of
37.5~K (see inset of fig.~\ref{fig-hc2}a). In sample B we find
$T_c$ = 38.3~K, $\Delta~T_c$ = 0.3~K and a "bulk $T_c$" of 38.2~K.
A simple analysis \cite{Ang91} of the slope of the magnetic moment
after reversing the applied field demonstrates that the size of
the domain, in which the supercurrents flow without impedance, is
identical to the sample size. Furthermore, a comparison of the
calculated and the measured magnetization in the Meissner regime
indicates a superconducting volume fraction of about 100~\%. The
further evaluation of the mixed state parameters did not show
significant differences between these two crystals.

The measurements of the magnetic moment were carried out in the
1~T and in an 8~T (SHE) SQUID magnetometer (for details, cf.
\cite{Sau98}). Fig.~\ref{fig-hc2}a shows the upper critical field
of MgB$_2$ for applied fields $H_a \| c$ ($H_{c2}^c$) and $H_a \|
ab$ ($H_{c2}^{ab}$). $H_{c2}(T)$ is determined either from the
onset of the superconducting signal in the $m(T)$ curve
("$T_c(H_a)$") or from the disappearance of the superconducting
signal in the $m(H_a)$ curve. The same results were obtained by
both methods. $H_{c2}^{ab}$ could be evaluated directly only below
8~T ($T >$ 21~K in this case). At lower temperatures the London
theory for the reversible magnetic moment $m_r$, i.e. $m_r \propto
\ln (H_{c2} / H_a)$, was used for the sake of simplicity to
extrapolate the experimental $m_r$ data to zero. The very small
magnetic moment in higher fields and the logarithmic behavior lead
to rather large uncertainties in the evaluation, which are
indicated by error bars in fig.~\ref{fig-hc2}a.

To obtain $H_{c2}^c(0)$, the data were fitted to the function
$H_{c2}(t) = H_{c2}(0)(1 - t^\alpha)^\beta$ with $µ_0H_{c2}^c(0)$
= 3.18~T ($t = T / T_c$, $T_c$ denotes the bulk transition
temperature; $\alpha$, $\beta$ and $H_{c2}(0)$ are fit
parameters). The initial slope of the upper critical field ($k =
µ_0 [\partial H_{c2}/\partial T]_{T_c})$ is found to be -0.112 T/K
near $T_c$, thus $µ_0H_{c2}^c(0) / (k T_c$) = -0.75. This is close
to the weak coupling BCS result ($\cong$ -0.73 in the clean limit,
-0.69 would correspond to the dirty limit \cite{Hel66}).
$H_{c2}(T)$ is not very sensitive to the coupling strength and the
above result, -0.75, is actually close to that of the strong
coupling superconductor Pb \cite{Car90}, of course without
considering anisotropy effects \cite{Haa01,Pit92}. If we apply the
same procedure to $H_a \| ab$, we obtain $µ_0H_{c2}^{ab}(0)$ =
15~T. However, the slope was determined in this case from the
linear region above 1~T ($k$ = -0.55~T/K) and the corresponding
extrapolated "$T_c$" of 36.1~K, because a strong positive
curvature of $H_{c2}^{ab}(T)$ is observed in the vicinity of $T_c$
(cf. the inset of fig.~\ref{fig-hc2}a), which represents a well
known feature of anisotropic superconductors, e.g. of the high
$T_c's$, but also of conventional superconductors, such as Nb. The
Fermi surface and the electron phonon coupling are usually held
responsible for these anisotropic properties. According to
\cite{Pit92}, the high temperature end of $H_{c2}^{ab}(T)$ can be
obtained by an anisotropic Fermi surface, but not by the coupling
alone. In Nb, e.g., $H_{c2}(T)$ is well explained by an
anisotropic Fermi velocity and an anisotropic electron phonon
coupling \cite{Web91}. A similar model could possibly apply in the
case of MgB$_2$, but alternative theories (e.g. a two band model
\cite{Shu01}) are also under discussion.

The upper critical field anisotropy $\gamma = H_{c2}^{ab} /
H_{c2}^c$ is shown in Fig.~\ref{fig-hc2}b (full squares). It
increases from about 1 near $T_c$ to 4.2 at 22~K, in qualitative
agreement with previous results \cite{Ang01,Sol01,Wel02}. The open
squares refer to results from torque measurements taken in a 9~T
(Quantum Design) PPMS system. In this case, the angular dependence
of the reversible torque is fitted to the anisotropic London
theory with three fit parameters $\gamma$, $H_{c2}^c$ and
$\lambda_{ab}$ (cf. \cite{Zec96}). A comparison of the latter two
parameters with results from the SQUID measurements shows the high
reliability of the evaluation. Note that this method does not lead
to the anisotropy of the upper critical field, but rather to that
of the magnetic penetration depth ($\lambda$), which can, in
general, deviate from $H_{c2}^{ab} / H_{c2}^c$. In MgB$_2$, both
seem to be the same, at least for $T \leq$ 30~K. The torque
indicates a small increase of $\gamma$ from 4.3 at 20~K to about
4.5 at 5~K leading to $\gamma(0) \cong 4.55$. However, we cannot
exclude some small systematic errors in the evaluation, because
(i) most of the recorded torque data refer to the irreversible
regime. Therefore, the reversible signal ($\tau$) has to be
calculated from the irreversible branches at increasing ($\tau_+$)
and decreasing ($\tau_-$) angles ($\tau = [\tau_+ + \tau_-] / 2$).
The difference between the two branches is rather small, but grows
at lower temperatures. (ii) The angular dependence of the
background signal varies with temperature and cannot be determined
exactly from measurements without a sample. However, different
data sets for the background do not change $\gamma$ at 5~K
significantly. Furthermore, the torque data were evaluated at
several magnetic fields (0.5 - 2~T), the differences in $\gamma$
were very small (2~\%). Based on the excellent agreement between
the SQUID and the torque data in the overlapping temperature
range, we assume that $\lambda_c / \lambda_{ab} = H_{c2}^{ab} /
H_{c2}^c$ for $T <$ 21~K, which allows us to calculate
$H_{c2}^{ab}$ in this temperature range (cf. the solid squares for
$H_{c2}^{ab}$ at $T < $ 21~K in fig.~\ref{fig-hc2}a). This leads
to $H_{c2}^{ab}$(0) = 14.5~T.

\begin{figure}
\centering \includegraphics[width = \columnwidth]{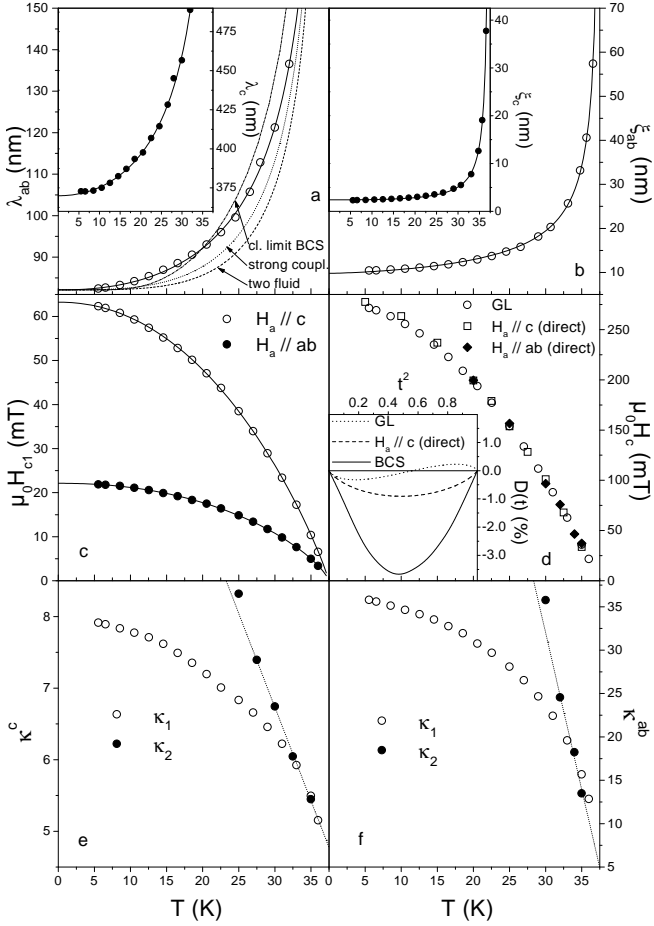}
\caption{ Temperature dependence of (a) the magnetic penetration
depths, (b) the coherence lengths, (c) the lower critical fields,
(d) the thermodynamic critical field and the deviation function
(inset) and (e, f) the GL parameters of sample A.} \label{fig-rev}
\end{figure}

\begin{figure}
\centering \includegraphics[width = \columnwidth]{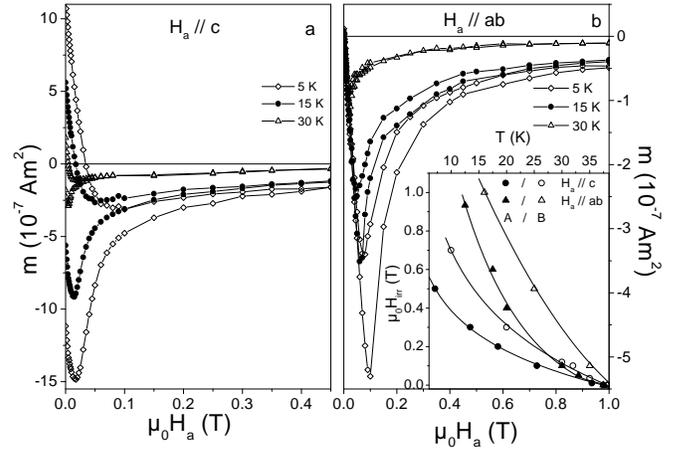}
\caption{Hysteresis loops of sample A for (a) $H_a \| c$ and (b)
$H_a \| ab$. Inset: Irreversibility line of samples A and B for
$H_a \| c$ and $H_a \| ab$.} \label{fig-irr}
\end{figure}

The further mixed state parameters can be calculated from the
London theory and some Ginzburg-Landau relations. For instance,
the magnetic penetration depth in the planes ($\lambda_{ab}$) is
obtained from $\partial M / \partial \ln (H_a) = \phi_0 / (8 \pi
\lambda_{ab}^2)$ for $H_a \| c$ ($M = m_r / volume$, $\phi_0 \cong
2.07 × 10^{-15}$ Vs). Since sample A shows a reversible
magnetization already at very small fields (cf.
fig.~\ref{fig-irr}), $\lambda_{ab}$ can be evaluated in the whole
temperature range from 5~K to $T_c$ (see fig.~\ref{fig-rev}a). A
fit of $\lambda_{ab}^{-2}(t)$ leads to $\lambda_{ab}(0)$ = 82~nm
and shows that the temperature dependence lies in between the
(clean limit) BCS \cite{Muh59} and a typical strong coupling model
\cite{Car90}. The deviation at lower temperatures indicates a
smaller energy gap - to - $T_c$ ratio than according to the BCS
theory, in agreement with other experiments (e.g.
\cite{Giu01,Sza01}), which can be explained by the two band
(assuming a small and a large gap \cite{Bou01}) as well as by the
anisotropic gap model \cite{Haa01}. Further work on more
comprehensive calculations including material dependent parameters
are currently under way. The penetration depth in c - direction is
obtained from $\lambda_c = \gamma \lambda_{ab}$, hence
$\lambda_c(0) = $ 370 nm. The evaluation of $\lambda_c$ from the
$m(T)$ measurements for $H_a \| ab$ confirms the anisotropy, but
is affected by comparatively large errors.

Further, the (GL) relation $µ_0H_{c2}^{ab} = \phi_0 / (2 \pi
\xi_{ab}^2$) gives access to the coherence length in the ab -
plane $\xi_{ab}$ and in the c - direction $\xi_c = \xi_{ab} /
\gamma$ (fig. \ref{fig-rev}b). Accordingly, $\xi_{ab}$(0) =
10.2~nm and $\xi_c(0) = 2.3~nm$.

The lower critical field $H_{c1}$ can be calculated from
$µ_0H_{c1}^x = [\phi_0 / (2 \pi \delta
\lambda_{ab}^2)]\cdot[\ln(\delta \lambda_{ab} / \xi_{ab}) + 0.5]$
($x$ = c and $\delta$ = 1 for $H_a \| c$ and $x$ = ab and $\delta
= \gamma$ for $H_a \| ab$), leading to $µ_0H_{c1}^c(0)$ = 63~mT
and $µ_0H_{c1}^{ab}(0)$ = 22~mT (fig.~\ref{fig-rev}c). A direct
experimental assessment of $H_{c1}$ is usually quite difficult,
because only the penetration field $H_p$, i.e. the field, at which
the first flux lines enter the sample, can be obtained, which
depends on the sample geometry \cite{Zel94}, the anisotropy and
the pinning force. We determined $H_p$ from measurements of the
trapped magnetic moment, i.e. by measuring the moment in zero
field after successively applying higher external fields and
searching for the first deviation from zero at $H_p$ \cite{Boh97}.
This procedure is still influenced by a finite critical current
density \cite{Kuz97} and by geometry effects. For example,
$µ_0H_p^c \cong$ 4~mT at 5~K (sample A) can be converted into
$H_{c1}(H_p)$ for a rectangular sample geometry \cite{Doy97},
which leads to $H_{c1} / H_p \cong $ 17.7 for $\gamma = 4.5$, i.e.
$\mu_0H_{c1}^c$(5~K) $\cong$ 70~mT, but overestimates $H_{c1}^c$
because the critical current density is not taken into account,
and is not too far away from the calculated result of
fig.~\ref{fig-rev}c (62~mT at 5~K).

Furthermore, the thermodynamic critical field is calculated from
the GL relation $\mu_0H_c = \phi_0 / (\sqrt{8} \pi \lambda_{ab}
\xi_{ab})$ and found to be 0.28~T at 0~K. Because $\Delta f =
\mu_0 H_c^2 / 2$ (condensation energy), it can also be obtained by
integrating the reversible magnetization $M(H_a)$, i.e. $\Delta f
= \mu_0 \int_0^{H_{c2}} M dH_a$. The reversible magnetic moment is
either calculated from the irreversible branches of the
magnetization in increasing ($m_+$) and decreasing ($m_-$) fields
in the fully penetrated state, $m_r = (m_+ + m_-) / 2$ (cf.
fig.~\ref{fig-irr}) or directly measured. The results of the
numerical integration are shown in fig.~\ref{fig-rev}d and denoted
by $H_a \| c$ - direct and $H_a \| ab$ - direct, respectively. A
comparison with the GL results indicates that the London model for
the magnetic penetration depth and the GL relations for $H_{c2}$
and $\xi$ represent excellent solutions for MgB$_2$. The maximum
difference at low temperatures is less than  2~\%. To check the
influence of uncertainties near $H_p$ in the direct evaluation
(geometrical barrier, flux pinning), we replace $M(H_a)$ at 5~K by
the separately measured Meissner slope at $0 \leq H_a \leq
H_{c1}(1-D)$ and by a simple logarithmic behavior at $H_{c1}(1-D)
\leq H_a \leq H_{c1}$, i.e. we simulate the behavior of an
ellipsoidal sample, where the "effective demagnetization factor"
$D$ is determined from $M(H_a) = - H_a / (1 - D)$ in the Meissner
regime. This procedure reduces $H_c$ at 5~K and brings the above
difference to almost zero.

The deviation function, $D(t) = [H_c(t) / H_c(0)] - [1-t^2]$,
describing the deviation of $H_c(t)$ from the parabolic behavior
(two fluid model) and indicating the coupling strength in a
conventional superconductor, is shown in the inset of
fig.~\ref{fig-rev}d. The maximum of -0.3 - -0.9~\% lies in between
the weak ($\sim$ -3.5~\%) and the strong coupling result ($\sim$
+2.5~\% for Pb). Although we have to consider evaluation errors,
the results indicate a clear deviation from the weak coupling
model, even if we consider the anisotropy (cf. \cite{Haa01}), and
is consistent with other experiments (e.g. \cite{Bud01,Qui02}).

The GL parameter $\kappa = \lambda / \xi$ is defined at $T = T_c$.
At lower temperatures the Maki parameters \cite{Mak64} $\kappa_1 =
H_{c2} / (\sqrt{2}H_c)$ and $\kappa_2 = [0.5 + 0.43 / (\partial M
/ \partial H_a)_{H_{c2}} - 0.43 D]^{1/2}$ can be used with
$\kappa_1(T_c) = \kappa_2(T_c) = \kappa$. $\kappa_1$ ($ = \lambda
/ \xi$ in the GL model) is shown in fig.~\ref{fig-rev}e for $H_a
\| c$. Linear extrapolations lead to $\kappa_1^c(0) = 8.1$ and
$\kappa_1^c(T_c) = 4.7$. The ratio $\kappa_1^c(0) /
\kappa_1^c(T_c) $ is 1.72 and considerably larger than the BCS
value (1.26 in the clean and 1.20 in the dirty limit
\cite{Hel66}), but this is not unexpected considering stronger
coupling \cite{Car90} and anisotropy. $\kappa_2$ depends on the
slope of $M$ near $H_{c2}$ and allows a precise determination of
$\kappa^c$, which is again found to be 4.7 from a linear
extrapolation to $T_c$. For $H_a \| ab$, we get $\kappa_1^{ab} =
\gamma \kappa_1^c$ and therefore $\kappa_1^{ab}(0) = 37.1$ and
$\kappa_1^{ab}(T_c) = 4.7$. The errors in $\kappa_2^{ab}$ are
relatively large in this case because of the very small slope of
the magnetization near $H_{c2}$, but the extrapolation leads to
$\kappa^{ab} = 5$, very close to $\kappa^c$.

At last, we turn to the irreversible properties of the MgB$_2$
single crystals. Hysteresis curves recorded at different
temperatures are presented in fig.~\ref{fig-irr} for $H_a \| c$
and $H_a \| ab$. They demonstrate the excellent crystal quality by
the small hysteresis and the low irreversibility fields in both
directions. Note that all data points presented in
fig.~\ref{fig-irr} were measured in the fully penetrated state.
According to the Bean model \cite{Bea62} ($J_c$ is assumed to be
constant), the critical current density in the planes can be
calculated from the irreversible magnetic moment ($m_i = [m_+ -
m_-] / 2$). For rectangular samples we use $J_c(B) = \{m_i(B) /
\Omega\}\{4/[b(1-b/3a)]\}$ (sample volume: $\Omega = a \cdot b
\cdot c$), and get $1.4 × 10^9$~Am$^{-2}$ at 5~K in the remnant
state for both samples. To obtain the irreversibility line, the
onsets of a difference between the field cooled and the zero field
cooled $m(T)$ measurement were evaluated. The results of
fig.~\ref{fig-irr} (inset) show that the irreversibility line is
very low for both field directions.

\begin{table}
\caption{\label{tab1} Summary of mixed state parameters for
MgB$_2$.}
\begin{ruledtabular}
\begin{tabular}{|l|r||l|r||l|r|}
$\mu_0H_{c2}^c(0)$&3.18~T&$\mu_0H_{c2}^{ab}(0)$&14.5~T&$T_c$&38~K\\
$\mu_0H_{c1}^c(0)$&63~mT&$\mu_0H_{c1}^{ab}(0)$&22~mT&$\mu_0H_c(0)$&0.28~T\\
$\lambda_c(0)$&370~nm&$\lambda_{ab}(0)$&82~nm&$\gamma(0)$&4.6\\
$\xi_c(0)$&2.3~nm&$\xi_{ab}(0)$&10.2~nm&$\gamma(T_c)$&1\\
$\kappa_1^c(0)$&8.1&$\kappa_1^{ab}(0)$&37.1&$\kappa(T_c)$&4.7\\
\end{tabular}
\end{ruledtabular}
\end{table}

In summary, we presented measurements of the magnetic moments in
single crystalline MgB$_2$ for fields $H_a \| c$ and $H_a \| ab$,
and the subsequent evaluation of the basic mixed state parameters.
The most important results are summarized in table~\ref{tab1}. The
general consistency of the data set, which is documented nicely,
e.g., by the results on the thermodynamic critical field, suggests
that the standard theoretical description can be employed in
MgB$_2$. The data indicate that MgB$_2$ is a low - $\kappa$ type
II superconductor in the clean limit with an intermediate electron
phonon coupling strength (cf. also \cite{Cho01a,Cho01b}), but a
very large anisotropy.

We wish to thank F. M. Sauerzopf for useful discussions and H.
Hartmann for technical assistance. This work was supported in part
by the Austrian Science Foundation (FWF project 14422), the
Austrian Exchange Service (OEAD 27/2000), the European Commission
(program ICA1-CT-2000-70018, Centre of Excellence CELDIS), the TMR
Network SUPERCURRENT and the Swiss National Science Foundation.

\end{document}